\documentstyle[amssymb,pra,aps]{revtex}

\newcommand{\be}{\begin{equation}}
\newcommand{\ee}{\end{equation}}
\newcommand{\br}{\begin{eqnarray}}
\newcommand{\er}{\end{eqnarray}}

\newcommand{\bd}{\begin{displaymath}}
\newcommand{\ed}{\end{displaymath}}

\newcommand{\bfig}{\begin{figure}}
\newcommand{\efig}{\end{figure}}

\def\3cdot{\cdot \cdot \cdot}

\def\g{\gamma}

\def\o{\omega}

\def\om0{\omega _0}
\def\Om0{\Omega _0}

\def\text#1{{\rm{#1}}}

\def\->{\rightarrow}
\def\=>{\Rightarrow}
\def\-->{\longrightarrow}
\def\==>{\Longrightarrow}

\def\pr{^\prime}

\def\pr2{^{\prime\prime}}

\def\bfig{\begin{figure}}
\def\efig{\end{figure}}

\begin{document}
\draft
\title{Continuous pumping and control of mesoscopic superposition state in a lossy
QED cavity}
\author{M. C. de Oliveira$^1$\thanks{%
E-mail: marcos@physics.uq.edu.au}, M. H. Y. Moussa$^2$\thanks{%
E-mail: miled@power.ufscar.br}, and S. S. Mizrahi$^2$\thanks{%
E-mail: salomon@power.ufscar.br},}
\address{$^1$Department of Physics, University of Queensland,\\
QLD 4072, Brisbane, Australia. \\
$^{2}$Departamento de F\'{\i }sica, CCET, Universidade Federal de S\~{a}o
Carlos,\\
Rod. Washington Luiz Km 235, S\~ao Carlos, 13565-905, SP, Brazil.}
\date{\today}
\maketitle
\begin{abstract}
Here we consider the continuous pumping of a dissipative QED cavity and derive
the time-dependent density operator of the cavity field prepared initially as a
superposition of mesoscopic coherent states. The control of the coherence of
this superposition is analyzed considering the injection of a beam of two-level
Rydberg atoms through the cavity. Our treatment is compared to other
approaches.
\end{abstract}


%

\section{Introduction}

%
Along the last two decades a consensus has been established about the
importance of the effects of the environment on a macroscopic system
to explain the non-observation of superposition of quantum states \cite
{sch,vonneumann,wignerm}. The formal treatment of a non-isolated macroscopic
quantum {\em system of interest} assumes a unitary evolution of the
whole system composed by
{\em system of interest} + {\em environment} + (possibly) {\em measurement
apparatus}. In the dynamical process called {\em decoherence, }the
environment drives the macroscopic superposition state into a statistical
mixture in a very short time, as compared to the relaxation time \cite{zurek}.
As a matter of fact, even microscopic systems suffer from the effects of the
environment since they are not perfectly isolated, however less drastically.

The construction of mesoscopic superposition states of the electromagnetic
field (EMF) in a cavity ({\em cat states}) has attracted attention due to
the experimental observation \cite {harochept,wineland} of its very short
lifetime as a superposition. The most recent proposals for preparing mesoscopic
superposition states relies on strategies whose aim consists in keeping them
in a high degree of purity, by precluding the noise coming from the
reservoir, in order to delay the decoherence process. One proposal for
creating and sustaining cat states is based on the confinement of an EMF in
a superconducting cavity \cite{haroche}. In \cite
{raibruhaprl,brune,davidovich} the authors proposed and achieved an
experiment consisting of the preparation of a cat state (a superposition
of coherent states in the microwave region) in a Fabry-Perot open
superconducting cavity of high quality factor then measuring its
decoherence time using the interaction with a beam of two-level
Rydberg atoms with an electromagnetic field going through the cavity
\cite{paulo}. In such an experiment energy and information loss
are unavoidable, not permitting the existence of the cat state for a
sufficiently long time, so constituting a drawback for its use for
technological purposes; the cavity field ends up in a thermal state \cite
{brune}.

Other proposals for suppressing the action of the environment on the coherence
of mesoscopic superposition field states have been presented in the last years.
One of them considers a stroboscopic feedback scheme \cite
{milburn,vitali,tombesi}: a stream of two-level atoms, all prepared in the same
state,{\bf \ }cross a cavity where each atom interacts dispersively with the
field. If the delay time between sequential atoms is short enough, all atoms
are to be detected in the same state, e.g., the excited state. This is an
indirect measure of the coherence and parity of the initial field superposition
state. However, when a cavity photon is lost{\bf ,} the following atom will be
detected in the ground state. When this event occurs a subsequent atom prepared
in the excited state should be sent to interact in resonance with the field in
order to compensate the lost energy and phase, so hopping to restore its
original state by absorbing a photon from the atom. This procedure needs a full
mastering of the atom-field interactions by the experimenter. However, due to
the poor efficiency of the atomic detectors, the stroboscopic feedback scheme
looses its full reliability.

In order to reduce the velocity of the decoherence process of a cat state in a
lossy QED cavity (hereby referred as C), in this paper another practical
strategy is proposed. It considers the continuous action of a classical pumping
field - a single mode microwave signal - in C, during the running time of the
experiment. We begin by showing that under the action of pumping and at
temperature $T=0$K, an arbitrary initial state of the field in C goes
asymptotically to a coherent state. The pumping action compensates the energy
lost to the environment, but not the initial available information about the
state (interference of probability amplitudes or coherence) as it is not
sensitive to the phase information of the field state. This can only be
achieved with the combined action of pumping together with the injection of
Rydberg atoms through the cavity, which permits sustaining the energy of the
field and reconstructing the initial coherence. This process can also use
feedback atoms, as proposed in \cite{milburn,vitali}, to guarantee a full
efficiency for maintaining the initial cat state. Another important question
raised in the present paper: For an open system constantly fed by an external
source how does evolve the decoherence and relaxation process?

This paper is organized as follows: In Section II we review the mechanism for
generating superposition field states in superconducting cavities. Section III
is devoted to obtain of the Heisenberg equations for the field operators that
govern the evolution of the continuously pumped quantum state. In Section IV we
discuss how a cat state is generated in a cavity and how it evolves when the
action of the combined pumping field plus environment proceeds. Section V is
dedicated to the study of the decoherence process of the cavity state. In
Section VI we propose a strategy which combines the action of atoms and pumping
to restore the initial superposition of the field state and finally in Section
VII we present a summary of this work.
%
\section{Generation of Schr\"odinger `cat' states}
%
The experimental apparatus for the generation of field superposition states
consist of a beam of Rydberg atoms crossing three cavities, R$_1$, R$_2$ and
C. R$_1$ and R$_2$ are low quality cavities (Ramsey zones), but C is a high-%
{\em Q} superconducting cavity, where a coherent state was previously
injected by a microwave source. The atoms are initially prepared in circular
states of quantum principal number of the order of 50 which are well
designed for these experiments since their life time is over $3\times
10^{-2} $s \cite{brune}.

The usual method of Ramsey interferometry consists in injecting classical
fields in the Ramsey zones R$_1$ and R$_2$ during the interaction time with
the atoms{\bf \ }\cite{haroche}. The transition between two nearly
orthogonal atomic states, $\left| e\right\rangle $ (excited) and $\left|
g\right\rangle $ (ground), is resonant with the R$_1$ and R$_2$ fields, and
the transition strength is set by selecting the velocity of the atom, which
suffers a rotation in the space spanned by state vectors
$\left\{ \left| e\right\rangle ,\left| g\right\rangle \right\} $.

The experiment begins by preparing the Rydberg atom in state $\left|
e\right\rangle $, which is then rotated in R$_1$ to the superposition state
\begin{equation}
\left| \Psi _a\right\rangle =\frac 1{\sqrt{2}}\left( \left| e\right\rangle
+\left| g\right\rangle \right) .  \label{a1}
\end{equation}
Subsequently the atom interacts with the field in $C$, whose dynamics is
described by the Jaynes-Cummings Hamiltonian
\begin{equation}
H=\hbar \omega a^{\dagger }a+\frac 12\hbar \omega _0\sigma _z+\hbar \kappa
\left( a\sigma ^{+}+a^{\dagger }\sigma ^{-}\right)  \label{a2}
\end{equation}
where $\sigma _z\equiv \left| e\right\rangle \left\langle e\right| -\left|
g\right\rangle \left\langle g\right| $, $\sigma ^{+}\equiv \left|
e\right\rangle \left\langle g\right| $ and $\sigma ^{-}\equiv \left|
g\right\rangle \left\langle e\right| $ are atomic pseudo-spin operators, $a$
($a^{\dagger }$) is the annihilation (creation) operator for the field mode
of frequency $\omega $ in $C$, $\kappa $ is the atom-field coupling
constant and $\omega _0$ is the atomic transition frequency.

The cavity $C$ is tuned near resonance with the atomic transition
frequency $\o _i$, between
states $\left| e\right\rangle $ and $\left| i\right\rangle $, where
$\left| i\right\rangle $ is a reference state with energy level
above that of $\left| e\right\rangle $. The transition frequency
$\omega _{i}$ is distinct of any other one involving the
state $\left| g\right\rangle $. The mode geometry inside the cavity is such
that the intensity of the field increases and decreases smoothly along the
atomic trajectory inside $C$. For sufficiently slow atoms and for
sufficiently large detuning between $\omega $ and $\omega _{i}$, the
atom-field evolution is adiabatic and no photonic absorption or emission
occurs \cite{haroche,paulo}. However, dispersive effects are very important
- the atom crossing $C$ in the state $\left| e\right\rangle $ induces a
phase shift in the cavity field which can be adjusted by a suited selection
of the atomic velocity ($\sim 100$ m/s). For a phase shift $\pi $, a
coherent field $\left| \alpha \right\rangle $ in $C$ is turned into $\left|
-\alpha \right\rangle $.
On the other hand, an atom in state $\left| g\right\rangle $ does not
introduce any phase shift on the cavity field. Therefore, an atom in
state (\ref{a1}) crossing $C$ will lead the system{\it \ C} + {\it atom}
in the correlated state
\begin{equation}
\frac{1}{\sqrt{2}}\left( \left| e\right\rangle +\left| g\right\rangle
\right) \otimes \left| \alpha \right\rangle \rightarrow \frac{1}{\sqrt{2}}%
\left( \left| e\right\rangle \otimes \left| -\alpha \right\rangle +\left|
g\right\rangle \otimes \left| \alpha \right\rangle \right) .  \label{a3}
\end{equation}
The atom crosses the cavity in a time of order of 10$^{-4}$ s, which is well
bellow the relaxation time of the field inside $C$ (typically 10$^{-3}$-10$%
^{-2}$s for Niobium superconducting cavities) and bellow the atomic
spontaneous emission time (3$\times 10^{-2}$s) \cite{paulo}.

When the atom is submitted to a second $\pi /2$ pulse, in R$_{2}$, the total
state will be transformed as
\begin{equation}
\frac{1}{\sqrt{2}}\left( \left| e\right\rangle \otimes \left| -\alpha
\right\rangle +\left| g\right\rangle \otimes \left| \alpha \right\rangle
\right) \rightarrow \frac{1}{\sqrt{2}}\left[ \left| e\right\rangle \otimes
\frac{1}{\sqrt{2}}\left( \left| -\alpha \right\rangle -\left| \alpha
\right\rangle \right) +\left| g\right\rangle \otimes \frac{1}{\sqrt{2}}%
\left( \left| -\alpha \right\rangle +\left| \alpha \right\rangle \right)
\right] .  \label{a4}
\end{equation}
Therefore, if the atom is detected in the state $\left| g\right\rangle $ or $%
\left| e\right\rangle $ the field in $C$ will be projected to the state
\begin{equation}
\left| \Psi _{c}\right\rangle =\frac{1}{N}\left( \left| \alpha \right\rangle
+\cos \varphi \left| -\alpha \right\rangle \right) ,  \label{a5}
\end{equation}
with $\varphi =0$ ($\pi $) if the atom is detected in the state $\left|
g\right\rangle $ ($\left| e\right\rangle $), $N=\sqrt{2\left( 1+\cos \varphi %
\mbox{e}^{-2\left| \alpha \right| ^{2}}\right) }$ is the%
normalization constant and the density operator for the superposition state%
{\em \ }(\ref{a5}) is given by
\begin{equation}
\rho _{C}=\left| \Psi _{c}\right\rangle \left\langle \Psi _{c}\right| =\frac{%
1}{N^{2}}\left\{ |\alpha ><\alpha |+|-\alpha ><-\alpha |+\cos \varphi \left(
\left| \alpha \right\rangle \left\langle -\alpha \right| +\left| -\alpha
\right\rangle \left\langle \alpha \right| \right) \right\} .  \label{a6}
\end{equation}
When such a state is produced inside the cavity, the presence of
dissipative effects alters its free evolution, introducing an amplitude
damping as well as a coherence loss term. At temperature $T=0$K the density
operator (\ref{a6}) evolves according to
\begin{eqnarray}
\rho _{C}(t) &=&\frac{1}{N^{2}}\left\{ |\alpha \mbox{e}^{-\gamma
t/2}><\alpha \mbox{e}^{-\gamma t/2}|+|-\alpha \mbox{e}^{-\gamma
t/2}><-\alpha \mbox{e}^{-\gamma t/2}|+\cos \varphi \mbox{e}^{-2|\alpha
|^{2}(1-\mbox {e}^{-\gamma t})}\right.  \nonumber \\
&&\left. \times \left[ |-\alpha \mbox{e}^{-\gamma t/2}><\alpha \mbox{e}%
^{-\gamma t/2}|+|\alpha \mbox{e}^{-\gamma t/2}><-\alpha \mbox{e}^{-\gamma
t/2}|\right] \right\} ,  \label{a11}
\end{eqnarray}
where two characteristic times are involved. The first one, the {\em %
decoherence time} is the time in which the pure state Eq.(\ref{a11}) is
turned into a statistical mixture
\begin{equation}
\rho _{C}(t)\approx \frac{1}{2}\left\{ |\alpha \mbox{e}^{-\gamma
t/2}><\alpha \mbox{e}^{-\gamma t/2}|+|-\alpha \mbox{e}^{-\gamma
t/2}><-\alpha \mbox{e}^{-\gamma t/2}|\right\} ;  \label{a8}
\end{equation}
the other one is the {\em damping }or{\em \ relaxation time} of the field, $%
t_{c} $ =$\gamma ^{-1}$, being the characteristic time when the energy
dissipation becomes important, driving the field asymptotically to a
vacuum state. The decoherence phenomenon is characterized by the factor $%
\exp \left[ -2\left| \alpha \right| ^{2}\left( 1-\mbox{e}^{-\gamma t}\right)
\right] $, and for short times, $\gamma t\ll 1$, it turns to be $\exp \left[
-2\left| \alpha \right| \gamma t\right] $. The {\em decoherence time} $%
t_{d}=\left( 2\gamma \left| \alpha \right| ^{2}\right) ^{-1}$ will be called
for future reference, the {\em free decoherence time}.
%
\section{Theory of classical pumping of lossy cavities}
%
We are going to show how a stationary coherent field state is generated in
cavities by the action of continuous pumping and how this can change the
decoherence process due to the energy loss. In the experimental apparatus
discussed in the last section the pumping consists in maintaining the
microwave radiation in C during the experimental running time.

A single EM mode in C interacts with the reservoir modes, represented
by a vast number of harmonic oscillators \cite{louisell,miledsalcal},
accounting for the energy dissipation of the field in C. In the
rotating wave approximation the total Hamiltonian is
\begin{equation}
H=\hbar \omega _{0}a^{\dagger }a+\sum _{k}\hbar \omega _{k}b_{k}^{\dagger
}b_{k}+\hbar \sum _{k}\left( \lambda _{k}a^{\dagger }b_{k}+\lambda
_{k}^{\ast }ab_{k}^{\dagger }\right) +\hbar \left[ F\mbox{e}^{-i\omega
t}a^{\dagger }+F^{\ast }\mbox{e}^{i\omega t}a\right]   \label{b1}
\end{equation}
where $\omega _{0} $ is the mode frequency of the cavity, $\omega
_{k} $ is the frequency of the $k$-th mode of the reservoir, $\lambda _{k} $
is the field-reservoir coupling constant and $F $ is the coupling constant
between the cavity and pumping fields, proportional to the pumping field
amplitude. Operators {$a^{\dagger } $ ($a $)} and {$b_{k}^{\dagger } $
($b_{k} $)} are
the bosonic creation (annihilation) operators of the field mode and the
reservoir, respectively. Let us suppose that initially the quantum field and
reservoir are uncoupled,
\begin{equation}  \label{b2}
\left| \Psi _{T};t=0\right\rangle \equiv \left| \psi _{F}\right\rangle
\otimes \left| \phi _{R}\right\rangle ,
\end{equation}
where $\left| \psi _{F}\right\rangle $ is the state of the field and $\left|
\phi _{R}\right\rangle $ is the state of the reservoir.

The Heisenberg equations for {$a$} and {$b_k$} are given by
\begin{eqnarray}
\dot{a} &=&\frac 1{i\hbar }\left[ a,H\right] =-i\omega
_0a-i\sum\nolimits_k\lambda _kb_k-iF\mbox{e}^{-i\omega t},  \label{b3} \\
\dot{b}_k &=&\frac 1{i\hbar }\left[ b_k,H\right] =-i\omega _kb_k-i\lambda
_k^{*}a,  \label{b4}
\end{eqnarray}
and the formal solution to Eq. (\ref{b4}) is
\begin{equation}
b_k(t)=\mbox{e}^{-i\omega _kt}b_k(0)-i\lambda _k^{*}\int_0^ta(t^{\prime })%
\mbox{e}^{i\omega _k(t^{\prime }-t)}dt^{\prime }.  \label{b5}
\end{equation}
The rapid oscillation of the free field evolution can be eliminated by
introducing in Eq. (\ref{b3}) the operator of slow variation in time $%
A\equiv \mbox{e}^{-i\omega _0t}a,$ whose equation of motion is
\begin{equation}
\dot{A}=-i\sum\nolimits_k\lambda _kb_k\mbox{e}^{i\omega _0t}-iF\mbox{e}%
^{i\omega _0t}\mbox{e}^{-i\omega t}.  \label{b7}
\end{equation}
Substituting Eq. (\ref{b5}) into Eq. (\ref{b7}) we get an equation for $A$
only,
\begin{equation}
\dot{A}=-i\sum\nolimits_k\lambda _kb_k(0)\mbox{e}^{-i(\omega _k-\omega
_0)t}-\sum_k\left| \lambda _k\right| ^2\int_0^tA(t^{\prime })\mbox{e}%
^{-i\left( \omega _k-\omega _0\right) (t^{\prime }-t)}dt^{\prime }-iF\mbox{e}%
^{i\omega _0t}\mbox{e}^{-i\omega t}.  \label{b8}
\end{equation}
Using the Wigner-Weisskopf approximation \cite{louisell,miledsalcal} into
the above equation (see details of calculations in Ap. A) and after some
algebraic manipulation the solution of the Heisenberg equation for the
operator $a$ writes as
\begin{equation}
a(t)=u(t)a(0)+\sum_kv_k(t)b_k(0)+w(t),  \label{b9}
\end{equation}
where
\begin{equation}
u(t)=\mbox{e}^{-\frac \gamma 2t}\mbox{e}^{-i\omega _0t},  \label{b10}
\end{equation}
\begin{equation}
v_k(t)=-\lambda _k\mbox{e}^{-i\omega _kt}\frac{\left[ 1-\mbox{e}^{-\frac %
\gamma 2t}\mbox{e}^{i\left( \omega _k-\omega _0\right) t}\right] }{\omega
_0-\omega _k-i\frac \gamma 2},  \label{b11}
\end{equation}
and
\begin{equation}
w(t)=F\mbox{e}^{-i\omega t}\frac{\left[ 1-\mbox{e}^{-\frac \gamma 2t}\mbox{e}%
^{i\left( \omega -\omega _0\right) t}\right] }{\omega -\omega _0+i\frac %
\gamma 2},  \label{b12}
\end{equation}
$\gamma $ (defined in Ap. A) is the damping constant.
%
\subsection{Characteristic function and field state representation}
%
Any density operator can be spanned by the overcomplete basis
of the coherent states having associated a Glauber-Sudarshan $P$-distribution,
\begin{equation}
\rho (t)=\int d^{2}\gamma P(\gamma ;t)\left| \gamma \right\rangle
\left\langle \gamma \right| .  \label{c1}
\end{equation}
The normal ordered characteristic function (CF) associated to $\rho (t)$
is given by
\begin{equation}
\chi _{N}(\eta ,t)={\rm Tr}\left[ \rho (t)\;\mbox{e}^{\eta a^{\dagger }}%
\mbox{e}^{-\eta ^{*}a}\right] ={\rm Tr}\left[ \rho (0)\;\mbox{e}^{\eta
a^{\dagger }(t)}\mbox{e}^{-\eta ^{*}a(t)}\right] ,  \label{c2}
\end{equation}
where the term in the middle is written in the Schr\"{o}dinger
picture and the last one is in the Heisenberg picture. The $P$
Glauber-Sudarshan distribution \cite{glauber,milburnli} is related to the
normal ordered CF by a double Fourier transform (FT)
\begin{equation}
P(\gamma ;t)=\frac{1}{\pi ^{2}}\int d^{2}\eta \mbox{e}^{\gamma \eta
^{*}-\gamma ^{*}\eta }\chi _{N}(\eta ,t),  \label{c3}
\end{equation}
whereas the Wigner function \cite{wigner} is defined as a double Fourier
transform of the symmetric ordered CF by
\begin{equation}
W(\zeta ;t)=\frac{1}{\pi ^{2}}\int d^{2}\eta \mbox{e}^{\zeta \eta ^{*}-\zeta
^{*}\eta }\chi _{S}(\eta ,t).  \label{c4}
\end{equation}
Both CF's are related through
\begin{equation}
\chi _{S}(\eta ,t)\equiv {\rm Tr}\left[ \rho (t)\;\mbox{e}^{\eta a^{\dagger
}-\eta ^{*}a}\right] =\mbox{e}^{-\frac{1}{2}\left| \eta \right| ^{2}}\chi
_{N}(\eta ,t).  \label{c5}
\end{equation}
Substituting Eq. (\ref{c5}) into Eq. (\ref{c4}) and using the inverse FT of
Eq. (\ref{c3}), we relate the Wigner function to the P-distribution as
\begin{equation}
W(\zeta ;t)=\frac{2}{\pi }\int d^{2}\gamma \mbox{e}^{-2\left| \zeta -\gamma
\right| ^{2}}P(\gamma ,t).  \label{c6}
\end{equation}
The symmetric ordered CF associated to the $\omega _{0}$ mode in cavity $C$
is given, in the Heisenberg picture, by
\begin{equation}
\chi _{S}^{F}(\eta ,t)={\rm Tr}_{F+R}\left[ \rho _{F+R}(0)\;\mbox{e}^{\eta
a^{\dagger }(t)-\eta ^{*}a(t)}\right] ,  \label{c7}
\end{equation}
where the trace operation runs over the field and reservoir coordinates and
the subsystems are assumed initially uncorrelated,{\bf \ }
\begin{equation}
\rho _{F+R}(0)=\rho _{F}(0)\otimes \rho _{R}(0).  \label{c8}
\end{equation}
Inserting operator (\ref{b9}) and its Hermitian conjugate into Eq. (\ref{c7}%
), the CF for the field writes
\begin{eqnarray}
\chi _{S}^{F}(\eta ,t) &=&{\rm Tr}_{F+R}\left\{ \rho _{F+R}(0)\;\exp \left[
\eta \left( u^{*}(t)a^{\dagger }+\sum_{k}v_{k}^{*}(t)b_{k}^{\dagger
}+w^{*}(t)\right) -h.c.\right] \right\}  \nonumber \\
&=&{\rm Tr}_{F+R}\left\{ \rho _{F+R}(0)\;\mbox{e}^{\eta
w^{*}(t)-n^{*}w(t)}\exp \left[ \eta u^{*}(t)a^{\dagger }-\eta
^{*}u(t)a\right] \exp \left[ \sum_{k}\left( \eta v_{k}^{*}(t)b_{k}^{\dagger
}-\eta ^{*}v_{k}(t)b_{k}\right) \right] \right\}  \nonumber \\
&=&\mbox{e}^{\eta w^{*}(t)-\eta ^{*}w(t)}{\rm Tr}_{F}\left\{ \rho
_{F}(0)\;\exp \left[ \eta u^{*}(t)a^{\dagger }-\eta ^{*}u(t)a\right] \right\}
\nonumber \\
&&\times {\rm Tr}_{R}\left\{ \rho _{R}(0)\;\exp \left[ \sum_{k}\left( \eta
v_{k}^{*}(t)b_{k}^{\dagger }-\eta ^{*}v_{k}(t)b_{k}\right) \right] \right\}
\nonumber \\
&=&\mbox{e}^{\eta w^{*}(t)-\eta ^{*}w(t)}\chi _{S}^{F}(\eta u^{*}(t),0){\rm %
Tr}_{R}\left\{ \rho _{R}(0)\;\exp \left[ \sum_{k}\left( \eta
v_{k}^{*}(t)b_{k}^{\dagger }-\eta ^{*}v_{k}(t)b_{k}\right) \right] \right\} ,
\label{c9}
\end{eqnarray}
with
\begin{equation}
\chi _{S}^{F}(\eta u^{*}(t),0)\equiv {{\rm Tr}_{F}\left[ \rho _{F}(0)\;%
\mbox{e}^{\eta u^{*}(t)a^{\dagger }-\eta ^{*}u(t)a}\right] .}  \label{c10}
\end{equation}
For a thermalized reservoir the state is given by
\begin{equation}
\rho _{R}(0)=\int \prod\nolimits_{k}d^{2}\beta _{k}\frac{1}{\pi \left\langle
n_{k}\right\rangle }\mbox{e}^{-\frac{\left| \beta _{k}\right| ^{2}}{%
\left\langle n_{k}\right\rangle }}\left| \beta _{k}\right\rangle
\left\langle \beta _{k}\right| ,  \label{c11}
\end{equation}
where $\left\langle n_{k}\right\rangle $ is the mean occupation number of
the $k$-th oscillator mode. So, Eq. (\ref{c10}) can be written as
\begin{equation}
\chi _{S}^{F}(\eta ,t)=\mbox{e}^{\eta w^{*}(t)-\eta ^{*}w(t)}\chi
_{S}^{F}(\eta u^{*}(t),0)\prod\nolimits_{k}\mbox{e}^{-\frac{1}{2}\left| \eta
\right| ^{2}\left| v_{k}(t)\right| ^{2}}\int d^{2}\beta _{k}\frac{1}{\pi
\left\langle n_{k}\right\rangle }\mbox{e}^{-\frac{\left| \beta _{k}\right|
^{2}}{\left\langle n_{k}\right\rangle }}\exp \left[ \eta v_{k}^{*}(t)\beta
_{k}^{\dagger }-\eta ^{*}v_{k}(t)\beta _{k}\right] .  \label{c12}
\end{equation}
The integral in Eq. (\ref{c12}) is easily solved whit the help of the
identity
\begin{equation}
\frac{1}{\pi }\int d^{2}\eta \mbox{e}^{-\mu \left| \eta \right| ^{2}+\lambda
\eta +\nu \eta ^{*}}=\frac{1}{\mu }\mbox{e}^{\frac{\lambda \nu }{\mu }%
},\;\;\left( \mbox {Re}\left( \mu \right) >0\right)  \label{c13}
\end{equation}
and the CF writes
\begin{eqnarray}
\chi _{S}^{F}(\eta ,t) &=&\mbox{e}^{\eta w^{*}(t)-\eta ^{*}w(t)}\chi
_{S}^{F}(\eta u^{*}(t),0)\prod\nolimits_{k}\mbox{e}^{-\left| \eta \right|
^{2}\left| v_{k}(t)\right| ^{2}\left( \frac{1}{2}+\left\langle
n_{k}\right\rangle \right) }  \nonumber \\
&=&\mbox{e}^{\eta w^{*}(t)-\eta ^{*}w(t)}\chi _{S}^{F}(\eta u^{*}(t),0)%
\mbox{e}^{-\left| \eta \right| ^{2}\sum_{k}\left| v_{k}(t)\right| ^{2}\left(
\frac{1}{2}+\left\langle n_{k}\right\rangle \right) }.  \label{c14}
\end{eqnarray}
Substituting the discrete sum by an integration we obtain after a minor
algebra \cite{miledsalcal}
\begin{equation}
\sum_{k}\left| v_{k}(t)\right| ^{2}\left( \frac{1}{2}+\left\langle
n_{k}\right\rangle \right) =\left( 1-\mbox{e}^{-\gamma t}\right) \left(
\frac{1}{2}+\bar{n}\right)  \label{c15}
\end{equation}
with $\bar{n}\equiv \left( \mbox{e}^{\beta \omega _{0}}-1\right) ^{-1}$, $%
\beta =\left( k_{B}T\right) ^{-1}$, where $k_{B}$ is the usual Boltzmann
constant and $T$ is the reservoir temperature. Substituting Eq. (\ref{c15})
into Eq. (\ref{c14}) we obtain
\begin{equation}
\chi _{S}^{F}(\eta ,t)=\chi _{S}^{F}(\eta u^{*}(t),0)\mbox{e}^{\eta
w^{*}(t)-\eta ^{*}w(t)}\mbox{e}^{-\left| \eta \right| ^{2}\left( 1-\mbox{e}%
^{-\gamma t}\right) \left( \frac{1}{2}+\bar{n}\right) }.  \label{c16}
\end{equation}
For a reservoir at $T=0$K, $\bar{n}=0$,{\bf \ }the symmetrically ordered CF
becomes
\begin{eqnarray}
\chi _{S}^{F}(\eta ,t) &=&{{\rm Tr}_{F}\left[ \rho _{F}(0)\;\mbox{e}^{\eta
u^{*}(t)a^{\dagger }-\eta ^{*}u(t)a}\right] \mbox{e}^{\eta w^{*}(t)-\eta
^{*}w(t)}\mbox{e}^{-\frac{1}{2}\left| \eta \right| ^{2}\left( 1-\mbox{e}%
^{-\gamma t}\right) }}  \nonumber \\
&=&{{\rm Tr}_{F}\left[ \rho _{F}(0)\;\mbox{e}^{\eta u^{*}(t)a^{\dagger }}%
\mbox{e}^{-\eta ^{*}u(t)a}\right] \mbox{e}^{\eta w^{*}(t)-\eta ^{*}w(t)}%
\mbox{e}^{-\frac{1}{2}\left| \eta \right| ^{2}};}  \label{c17}
\end{eqnarray}
comparing the RHS of the second equality with the normal ordered CF,
Eq. (\ref{c5}), we identify the following relation
\begin{equation}
\chi _{N}^{F}(\eta ,t)=\chi _{N}^{F}(\eta u^{*}(t),0)\mbox{e}^{\eta
w^{*}(t)-\eta ^{*}w(t)}.  \label{c18}
\end{equation}
At this point it is important to emphasize that we have not mentioned yet the
initial state of the field inside the cavity. Eq. (\ref{c18}) allows one to
obtain the evolved density operator for an arbitrary initial state. The
dynamics of the system {\em cavity field} + {\em reservoir} correlates the
initial states of the subsystems entailing energy dissipation and loss of
coherence during evolution. In the next section we show that when the reservoir
is at $T=0$K, both, the cavity field and the reservoir states, remain
uncorrelated in the course of the evolution. In the absence of pumping, the
field state is called {\em dissipative coherent state}.
%
\section{Generation of states in the dissipative cavity}
%
\subsection{Coherent states}
%
Let us first consider the situation when the initial state of the field in
the cavity C is
\begin{equation}
\rho _C(0)=\left| \alpha \right\rangle \left\langle \alpha \right| ;
\label{d1}
\end{equation}
introducing it into the normal ordered CF Eq. (\ref{c2}) we have
\begin{eqnarray}
\chi _N^F(\eta u^{*}(t),0) &=&{{\rm Tr}_F\left[ \left| \alpha \right\rangle
\left\langle \alpha \right| \;\mbox{e}^{\eta u^{*}(t)a^{\dagger }}\mbox{e}%
^{-\eta ^{*}u(t)a}\right] }  \nonumber \\
&=&\left\langle \alpha \right| \mbox{e}^{\eta u^{*}(t)a^{\dagger }}\mbox{e}%
^{-\eta ^{*}u(t)a}\left| \alpha \right\rangle  \nonumber \\
&=&\mbox{e}^{\eta u^{*}(t)\alpha ^{*}-\eta ^{*}u(t)\alpha },  \label{d2}
\end{eqnarray}
and substituting into Eq. (\ref{c18}) one gets
\begin{equation}
\chi _N^F(\eta ,t)=\mbox{e}^{\eta \left[ u^{*}(t)\alpha ^{*}+w^{*}(t)\right]
-\eta ^{*}\left[ u(t)\alpha +w(t)\right] }.  \label{d3}
\end{equation}
However, since this result must be identical to the normal ordered CF
obtained in the Schr\"{o}dinger picture,{\bf \ }it follows that
\begin{equation}
\chi _N^F(\eta ,t)={{\rm Tr}_F\left[ \rho _F(t)\;\mbox{e}^{\eta a^{\dagger }}%
\mbox{e}^{-\eta ^{*}a}\right] =\left\langle \psi _F(t)\right| \mbox{e}^{\eta
a^{\dagger }}\mbox{e}^{-\eta ^{*}a}\left| \psi _F(t)\right\rangle ,}
\label{d4}
\end{equation}
with $\rho _F(t)=\left| \psi _F(t)\right\rangle \left\langle \psi
_F(t)\right| $. Then, if we compare the term at the LHS of the second
equality of Eq. (\ref{d4}) to Eq. (\ref{d3}), it can be directly verified
that one gets
\begin{equation}
\left| \psi _F(t)\right\rangle =\left| u(t)\alpha +w(t)\right\rangle ,
\label{d7}
\end{equation}
as a consequence of the disentanglement between the field and the reservoir
states, only at $T=0$K. Thus the density operator for the continuously
pumped field is given by
\begin{eqnarray}
\rho _F(t) &=&\left| u(t)\alpha +w(t)\right\rangle \left\langle u(t)\alpha
+w(t)\right|  \nonumber \\
&=&\left| \mbox{e}^{-\frac \gamma 2t}\mbox{e}^{-i\omega _0t}\alpha
+w(t)\right\rangle \left\langle \mbox{e}^{-\frac \gamma 2t}\mbox{e}%
^{-i\omega _0t}\alpha +w(t)\right| ,  \label{d8}
\end{eqnarray}
where
\begin{equation}
w(t)=F\mbox{e}^{-i\omega t}\frac{\left[ 1-\mbox{e}^{-\frac \gamma 2t}\mbox{e}%
^{i\left( \omega -\omega _0\right) t}\right] }{\omega -\omega _0+i\frac %
\gamma 2}.  \label{d9}
\end{equation}
By adjusting the pumping field in resonance with the cavity field ($\omega
=\omega _0$), we have
\begin{equation}
w(t)=-i\frac{2F}\gamma \mbox{e}^{-i\omega _0t}\left( 1-\mbox{e}^{-\frac %
\gamma 2t}\right) ,  \label{d10}
\end{equation}
and the density operator becomes
\begin{equation}
\rho _F(t)=\left| \mbox{e}^{-i\omega _0t}\left[ \mbox{e}^{-\frac \gamma 2%
t}\alpha -i\frac{2F}\gamma \left( 1-\mbox{e}^{-\frac \gamma 2t}\right)
\right] \right\rangle \left\langle \mbox{e}^{-i\omega _0t}\left[ \mbox{e}^{-%
\frac \gamma 2t}\alpha -i\frac{2F}\gamma \left( 1-\mbox{e}^{-\frac \gamma 2%
t}\right) \right] \right| .  \label{d11}
\end{equation}
Setting the relation between the system parameters, $F=i\alpha \gamma /2$,
all the terms multiplying $\exp (-\gamma t/2)$ cancel and the field in the
cavity remains coherent, oscillating at frequency $\omega _0$,
\begin{equation}
\rho _C(t)=\left| \mbox{e}^{-i\omega _0t}\alpha \right\rangle \left\langle %
\mbox{e}^{-i\omega _0t}\alpha \right| .  \label{d12}
\end{equation}
In this way despite the dissipative effect, the pumping action compensates
the lost energy, establishing the stationary coherent field state in the
cavity. This result is independent of the cavity quality factor $Q\equiv
\omega _0/\gamma $, showing that coherent fields are quite stable.

For the generation of another coherent state it is sufficient to adjust the
pumping field amplitude. Asymptotically the field state is stationary,
\begin{equation}
\lim_{t\rightarrow \infty }\rho _F(t)\approx \left| \mbox{e}^{-i\left(
\omega _0t+\frac \pi 2\right) }\frac{2F}\gamma \right\rangle \left\langle %
\mbox{e}^{-i\left( \omega _0t+\frac \pi 2\right) }\frac{2F}\gamma \right| ,
\label{d14}
\end{equation}
even if the field in the cavity is initially in the vacuum state
$\alpha =0$. This result shows how a lossy cavity fills up coherently when
pumped by a classical source of EM radiation \cite{charmichael}.
%
\subsection{Superposition state}
%
Let us consider now that the state (\ref{d12}) is sustained in the cavity and
that the experiment described in Sec. 2 is going on. With the pumping field
acting continuously we consider the adiabatic passage of a Rydberg atom across
the cavity $C$, {\it i.e.}, the time of flight of the atom is very small
compared to the relaxation time of the field. It is worth noting that if the
detuning between the atomic transition frequency $\omega _i$ and the cavity
field is sufficiently large, the atomic presence inside the cavity do not
changes considerably its frequency mode distribution. For an atom prepared
initially in the state $\left| e\right\rangle $ the density operator of the
system {\em atom+field} is written as
\begin{equation}
\rho _{F+A}=\rho _A\otimes \rho _F=\left| e\right\rangle \left\langle
e\right| \otimes \rho _F.  \label{e1}
\end{equation}
The resonant interaction of the atom with the field in $R_1$, rotates
the atomic state by a $\pi /2$,
\begin{equation}
\left| e\right\rangle \rightarrow \frac 1{\sqrt{2}}\left( \left|
e\right\rangle +\left| g\right\rangle \right) ,  \label{e2}
\end{equation}
and the joint density operator writes as
\begin{equation}
\rho _{F+A}=\frac 12\left( \left| e\right\rangle +\left| g\right\rangle
\right) \left( \left\langle e\right| +\left\langle g\right| \right) \otimes
\rho _F.  \label{e3}
\end{equation}
Due to the dispersive interaction of the atom with the field in $C$ the
joint state is given by \cite{davidovich}
\begin{equation}
\rho _{F+A}=\frac 12\left( \left| e\right\rangle \left\langle e\right| %
\mbox{e}^{-i\pi a^{\dagger }a}\rho _F\mbox{e}^{i\pi a^{\dagger }a}+\left|
g\right\rangle \left\langle g\right| \rho _F+\left| e\right\rangle
\left\langle g\right| \mbox{e}^{-i\pi a^{\dagger }a}\rho _F+\left|
g\right\rangle \left\langle e\right| \rho _F\mbox{e}^{i\pi a^{\dagger
}a}\right) ,  \label{e4}
\end{equation}
once the state $\left| e\right\rangle $ is always associated with the phase
shift operator $\exp (-i\pi a^{\dagger }a)$ in this experiment\cite
{brune,davidovich}. Then the outgoing atom passing through $R_2$ is
submitted to a new $\pi /2$ rotation and the joint state becomes
\begin{eqnarray}
\rho _{F+A} &=&\frac 14\left( \left| e\right\rangle +\left| g\right\rangle
\right) \left( \left\langle e\right| +\left\langle g\right| \right) \mbox{e}%
^{-i\pi a^{\dagger }a}\rho _F\mbox{e}^{i\pi a^{\dagger }a}+\left( -\left|
e\right\rangle +\left| g\right\rangle \right) \left( -\left\langle e\right|
+\left\langle g\right| \right) \rho _F  \nonumber \\
&&+\left( \left| e\right\rangle +\left| g\right\rangle \right) \left(
-\left\langle e\right| +\left\langle g\right| \right) \mbox{e}^{-i\pi
a^{\dagger }a}\rho _F+\left( -\left| e\right\rangle +\left| g\right\rangle
\right) \left( \left\langle e\right| +\left\langle g\right| \right) \rho _F%
\mbox{e}^{i\pi a^{\dagger }a}.
\end{eqnarray}
If the atom is detected in the state $\left| g\right\rangle $ or $\left|
e\right\rangle $, the field state collapses instantaneously to
\begin{equation}
\rho _F^{{g \choose {e}}
}=\frac 14\left[ \mbox{e}^{-i\pi a^{\dagger }a}\rho _F\mbox{e}^{i\pi
a^{\dagger }a}+\rho _F+\cos \varphi \left( \mbox{e}^{-i\pi a^{\dagger
}a}\rho _F+\rho _F\mbox{e}^{i\pi a^{\dagger }a}\right) \right] ,  \label{e6}
\end{equation}
where
\begin{equation}
\rho _F^g=\left\langle g\right| \rho _{F+A}\left| g\right\rangle ,\quad {\rm
{\ }}\rho _F^e=\left\langle e\right| \rho _{F+A}\left| e\right\rangle
\label{e7b}
\end{equation}
and $\varphi =0$ or $\pi $ depending on the atom being detected in state
$\left| g\right\rangle $ or $\left| e\right\rangle $, respectively. The
final state can be obtained from Eq. (\ref{e6}) when the initial state is
known; for example, if $\rho _F=\left| \alpha \right\rangle \left\langle \alpha
\right| $ is the initial state of the field in $C$, we have from Eq. (\ref
{d12}) ($\alpha $ containing the time-dependent phase $e^{-i\omega _0t}$)
\begin{equation}
\rho _F^{{g \choose {e}}
}=\frac 1{N^2}\left[ \left| \alpha \right\rangle \left\langle \alpha \right|
+\left| -\alpha \right\rangle \left\langle -\alpha \right| +\cos \varphi
\left( \left| -\alpha \right\rangle \left\langle \alpha \right| +\left|
\alpha \right\rangle \left\langle -\alpha \right| \right) \right] ,
\label{e8}
\end{equation}
with $N=\sqrt{2\left( 1+\cos \varphi \mbox{e}^{-2\left| \alpha \right|
^2}\right) }$. Then, immediately after the atomic detection, the collapsed
state of the field decoheres due to the effects of pumping and energy
dissipation.
Its explicit time dependence is obtained by first constructing the CF by
substituting Eq. (\ref{e8}) into Eq. (\ref{c18}),
\begin{eqnarray}
\chi _N(\eta ,t) &=&\frac 1{N^2}\left\{ \mbox{e}^{\eta \left( u^{*}(t)\alpha
^{*}+w^{*}(t)\right) -\eta ^{*}\left( u(t)\alpha +w(t)\right) }+\mbox{e}%
^{-\eta \left( u^{*}(t)\alpha ^{*}-w^{*}(t)\right) +\eta ^{*}\left(
u(t)\alpha -w(t)\right) }\right.  \nonumber \\
&&\left. \cos \varphi \mbox{e}^{-2\left| \alpha \right| ^2}\left[ \mbox{e}%
^{\eta \left( u^{*}(t)\alpha ^{*}+w^{*}(t)\right) +\eta ^{*}\left(
u(t)\alpha -w(t)\right) }+\mbox{e}^{-\eta \left( u^{*}(t)\alpha
^{*}-w^{*}(t)\right) -\eta ^{*}\left( u(t)\alpha +w(t)\right) }\right]
\right\} ,  \label{e9}
\end{eqnarray}
then by comparing, again, the expressions in both, the Schr\"{o}dinger
and Heisenberg pictures, we obtain the density operator for the field state
\begin{eqnarray}
\rho _F(t) &=&\frac 1{N^2}\left\{ \left| \mbox{e}^{-\frac \gamma 2t}\alpha
+w(t)\right\rangle \left\langle \mbox{e}^{-\frac \gamma 2t}\alpha
+w(t)\right| +\left| -\left( \mbox{e}^{-\frac \gamma 2t}\alpha -w(t)\right)
\right\rangle \left\langle -\left( \mbox{e}^{-\frac \gamma 2t}\alpha
-w(t)\right) \right| \right.  \nonumber \\
&&+\cos \varphi \mbox{e}^{-2\left| \alpha \right| ^2(1-\mbox{e}^{-\gamma
t})}\left[ \mbox{e}^{-\left[ \mbox{e}^{-\gamma t}\left( \alpha
^{*}w(t)-\alpha w^{*}(t)\right) \right] }\left| -\left( \mbox{e}^{-\frac %
\gamma 2t}\alpha -w(t)\right) \right\rangle \left\langle \mbox{e}^{-\frac %
\gamma 2t}\alpha +w(t)\right| \right.  \nonumber \\
&&\left. \left. +\mbox{e}^{\left[ \mbox{e}^{-\gamma t}\left( \alpha
^{*}w(t)-\alpha w^{*}(t)\right) \right] }\left| \mbox{e}^{-\frac \gamma 2%
t}\alpha +w(t)\right\rangle \left\langle -\left( \mbox{e}^{-\frac \gamma 2%
t}\alpha -w(t)\right) \right| \right] \right\} .  \label{h}
\end{eqnarray}
When the amplitude of the field is adjusted to $F=i\alpha \gamma /2$ we have
\begin{equation}
w(t)=\alpha \left( 1-\mbox{e}^{-\frac \gamma 2t}\right) ,  \label{e11}
\end{equation}
and
\begin{eqnarray}
\rho _F(t) &=&\frac 1{N^2}\left\{ \left| \alpha \right\rangle \left\langle
\alpha \right| +\left| \alpha \left( 1-2\mbox{e}^{-\frac \gamma 2t}\right)
\right\rangle \left\langle \alpha \left( 1-2\mbox{e}^{-\frac \gamma 2%
t}\right) \right| \right.  \nonumber \\
&&+\cos \varphi \mbox{e}^{-2\left| \alpha \right| ^2(1-\mbox{e}^{-\gamma
t})}\left[ \left| \alpha \left( 1-2\mbox{e}^{-\frac \gamma 2t}\right)
\right\rangle \left\langle \alpha \right| +\left| \alpha \right\rangle
\left\langle \alpha \left( 1-2\mbox{e}^{-\frac \gamma 2t}\right) \right|
\right] ,  \label{e12}
\end{eqnarray}
which shows the time evolution of the quantum state.
Asymptotically this state goes to the coherent state $\rho _F(t)=|\alpha
\rangle \langle \alpha |$, which acts as an attractor for other initial
quantum states. Independently of the initial amplitude $\alpha $, the
pumping field supplies energy continuously to the cavity, sustaining the
field in a pure coherent state, Eq. (\ref{d14}).
%
\section{Decoherence of a continuously pumped field}
%
Now we analyze the evolution of the density operator at times far from the
asymptotic regime, in which case we can observe the effects of the classical
pumping on the evolution of the state. From Eq. (\ref{h}) we observe that the
coherence terms (non-diagonal) are modified, in comparison to the pumping-free
decoherence, by a factor $\exp \left[ \pm \mbox{e}^{-\gamma t}\left( \alpha
^{*}w(t)-\alpha w^{*}(t)\right) \right] $. However, due to the oscillatory
character of these factors, the non-diagonal terms do not sustain the coherence
of the field which is continuously attenuated by the damping factor $\exp
\left[ -2\left| \alpha \right| ^2(1-\mbox{e}^{-\gamma t})\right] ,$ as shown in
Sec. 2.

The quantum characteristic of a field state can be visualized when
represented by a Wigner function \cite{wigner}, obtained from the Fourier
transform of the symmetrically ordered CF Eq. (\ref{c4}). The state given by
Eq. (\ref{h}) has as Wigner function
\begin{eqnarray}
W(\zeta ,t) &=&\frac{2}{N^{2}\pi }\left\{ \exp \left[ -2\left| \zeta -%
\mbox{e}^{-\gamma t/2}\alpha -w(t)\right| ^{2}\right] +\exp \left[ -2\left|
\zeta +\mbox{e}^{-\gamma t/2}\alpha -w(t)\right| ^{2}\right] \right.
\nonumber \\
&&\left. 2\cos \varphi \mbox{e}^{-2\left| \zeta -w(t)\right| ^{2}}\exp
\left[ -2\left| \alpha \right| ^{2}\left( 1-\mbox{e}^{-\gamma t}\right)
\right] \cos \left[ 4\mbox{e}^{-\gamma t/2}\mbox {Im}\left[ \left( \zeta
-w(t)\right) \alpha ^{*}\right] \right] \right\} ,  \label{f1}
\end{eqnarray}
where the first two exponential functions are Gaussians centered at $\left[ %
\mbox{e}^{-\gamma t/2}\alpha +w(t)\right] $ and $-\left[ \mbox{e}^{-\gamma
t/2}\alpha -w(t)\right] $ respectively, representing the two distinct
states, $\left| \mbox{e}^{-\frac{\gamma }{2}t}\alpha +w(t)\right\rangle $
and $\left| -\mbox{e}^{-\frac{\gamma }{2}t}\alpha +w(t)\right\rangle $. The
(third) coherence term is composed by three factors, a Gaussian centered at $%
w(t)$, a sinoid modulation, $\cos \left[ 4\mbox{e}^{-\gamma t/2}%
\mbox
{Im}\left[ \left( \zeta -w(t)\right) \alpha ^{*}\right] \right] $ and the
factor responsible for the decoherence, $\exp \left[ -2\left| \alpha \right|
^{2}\left( 1-\mbox{e}^{-\gamma t}\right) \right] $. The modulation and the
time of decoherence given by the last factor depend on the intensity of the
state. The larger is $\left| \alpha \right| ^{2}$ the faster is the
decoherence.

In Figs. 1-3 three configurations of the Wigner function (\ref{f1}) for $%
\left| \alpha \right| ^2=5$ are shown at three distinct times, $t=0$, $%
t=\gamma ^{-1}$ and $t\rightarrow \infty $. For $F=1$ we observe the
progressive evolution of the superposition state, driven continuously to a
stationary coherent state, Fig. 3, representing
\begin{equation}
\lim_{t\rightarrow \infty }W(\zeta ,t)= \frac{2}{\pi}
\exp \left[ -2\left| \zeta -\mbox{e}%
^{-i\left( \omega _0t+\frac \pi 2\right) }\frac{2F}\gamma \right| ^2\right] .
\label{f2}
\end{equation}
The coherence term is suppressed in a time shorter than the time of
relaxation of the state, still given by the free decoherence time $%
t_d=\left( 2\left| \alpha \right| ^2\gamma \right) ^{-1}$, being null the
effect of the pumping on the coherence terms.

The evolution of the superposition state shows the decoherence and
relaxation processes (loss of purity) as analyzed through the
linear entropy,
\begin{eqnarray}
S &=&{\rm Tr_{F}\left[ \rho _{F}(t)-\rho _{F}^{2}(t)\right] }  \nonumber \\
&=&1-\frac{2}{N^{4}}\left\{ 1+4\mbox{e}^{-2\left| \alpha \right| ^{2}}+%
\mbox{e}^{-4\left| \alpha \right| ^{2}\mbox{e}^{-\gamma t}}+\mbox{e}%
^{-4\left| \alpha \right| ^{2}\left( 1-\mbox{e}^{-\gamma t}\right) }\right.
\nonumber \\
&&\left. +\mbox{e}^{-4\left| \alpha \right| ^{2}}\mbox{e}^{-2\left| \alpha
\right| ^{2}\mbox{e}^{-\gamma t}}\cos \left[ 2\mbox{e}^{-\gamma t/2}%
\mbox
{Im}\left( w(t)\alpha ^{*}\right) \right] \right\} .  \label{f3}
\end{eqnarray}
In Fig. 4 we plotted $S$ against $\g t$ for $\left| \alpha \right| ^{2}=5$,
where the state is initially pure. As the decoherence goes on the state
evolves into a mixture, ${\rm Tr\rho ^{2}<1}$, and the entropy increases
meaning that there is a flux of information to the reservoir. Although the
pumping field is able to restore the energy lost by the cavity field, it is
not able to establish back the {\em information} of the original
superposition state encoded by the coherence terms. Certainly here the
process of decoherence is tied to the loss of energy of the field to the
reservoir; however there are situations where the information transfer does
not occur necessarily together with an energy transfer \cite{brune,meu3}. The
information transfer strongly depends on the phase relation of the
superposition of quantum states \cite{meu3}. Despite that reversible
subsystems can exhibit the decoherence and recoherence at constant mean
energy \cite{meu3}, this characteristic ceases to be true for irreversible
subsystems. Decoherence still occurs in a characteristic time, which is
dependent on the field relaxation time and the field energy, as shown in
Sec. (2). An open question remains: Is the information flow (decoherence)
always accompanied with an energy flow, or this is only valid for open
irreversible systems? We emphasize that the time irreversible character of
these models of reservoirs follows from the introduction of
approximations as the Wigner-Weisskopf and Markov.
%
\section{Atoms and pumping}
%
The attempt to sustain the field, against decoherence, in a superposition of
coherent states by using a classical pumping field is not effective because the
insertion of photons for compensating those lost to the reservoir is not phase
sensitive. The pumping is only sufficient to re-establish the energy lost to
the reservoir and not the original superposition state. Asymptotically only a
stationary coherent state is established in the cavity. However, the
maintenance of the superposition state could be possible if an additional
process accounting for re-establishing the original coherence is considered. In
the experiment proposed in \cite{raibruhaprl,brune}, once the superposition is
created in $C$, the field interacts with atoms sent sequentially through $C$.
The authors argue that this procedure refreshes the initial coherence. Here we
analyze the same process of sending atoms through the cavity, but with the
pumping field included.

At time $T,$ after the detection of the first atom,{\bf \ }the state of the
field in $C$ is given by $\rho _F(T)$, Eq. (\ref{h}); then a second atom is
released, going through the same interaction process as the former. After
crossing $R_1$, the second atom + $C$-field joint state is given by
\begin{equation}
\rho _{F+A_2}(T)=\frac 12\left( \left| e\right\rangle +\left| g\right\rangle
\right) _2\left( \left\langle e\right| +\left\langle g\right| \right)
_2\otimes \rho _F(T),  \label{g2}
\end{equation}
and the dispersive interaction in the $C$-field produces the entangled joint
state
\begin{eqnarray}
\rho _{F+A_2}(T) &=&\frac 12\left( \left| e\right\rangle \left\langle
e\right| _2\mbox{e}^{-i\pi a^{\dagger }a}\rho _F(T)\mbox{e}^{i\pi a^{\dagger
}a}+\left| g\right\rangle \left\langle g\right| _2\rho _F(T)+\left|
e\right\rangle \left\langle g\right| _2\mbox{e}^{-i\pi a^{\dagger }a}\rho
_F(T)\right.  \nonumber \\
&&\left. +\left| g\right\rangle \left\langle e\right| _2\rho _F(T)\mbox{e}%
^{i\pi a^{\dagger }a}\right) .  \label{g3}
\end{eqnarray}
After crossing the cavity $R_2$ the joint state suffers a new
transformation, becoming
\begin{eqnarray}
\rho _{F+A_2}(T) &=&\frac 14\left[ \left( \left| e\right\rangle +\left|
g\right\rangle \right) _2\left( \left\langle e\right| +\left\langle g\right|
\right) _2\mbox{e}^{-i\pi a^{\dagger }a}\rho _F(T)\mbox{e}^{i\pi a^{\dagger
}a}+\left( -\left| e\right\rangle +\left| g\right\rangle \right) _2\left(
-\left\langle e\right| +\left\langle g\right| \right) _2\rho _F(T)\right.
\nonumber \\
&&\left. +\left( \left| e\right\rangle +\left| g\right\rangle \right)
_2\left( -\left\langle e\right| +\left\langle g\right| \right) _2\mbox{e}%
^{-i\pi a^{\dagger }a}\rho _F(T)+\left( -\left| e\right\rangle +\left|
g\right\rangle \right) _2\left( \left\langle e\right| +\left\langle g\right|
\right) _2\rho _F(T)\mbox{e}^{i\pi a^{\dagger }a}\right] .  \nonumber \\
&&  \label{g4}
\end{eqnarray}
If the atom is detected in the $\left| g\right\rangle $ or $\left|
e\right\rangle $ state the field will collapse instantaneously to
\begin{equation}
\rho _F^{{g \choose {e}}
}(T)=\frac 14\left( \mbox{e}^{-i\pi a^{\dagger }a}\rho _F(T)\mbox{e}^{i\pi
a^{\dagger }a}+\rho _F(T)\pm \mbox{e}^{-i\pi a^{\dagger }a}\rho _F(T)\pm
\rho _F(T)\mbox{e}^{i\pi a^{\dagger }a}\right) ,  \label{g5}
\end{equation}
with the signal $+$ ($-$) standing for $\left| g\right\rangle $ ($\left|
e\right\rangle $).

Substituting Eq. (\ref{h}) for $\rho _{F}(T)$ in Eq. (\ref{g5}) we obtain the
conditional expression for $\rho _{F}^{{g \choose {e}}}(T)$.
In short, the probability for the second atom be detected in either
state $\left| g\right\rangle $ or $\left| e\right\rangle $ is given by
\begin{eqnarray}
P_{{g \choose {e}}}(T) &=&{{\rm Tr}_{F}\left[ \rho _{F}^{{g \choose {e}}
}(T)\right] }  \nonumber \\
&=&\frac{1}{2}\left( 1\pm \mbox {Re}\left\{ {{\rm Tr}\left[ \mbox{e}^{-i\pi
a^{\dagger }a}\rho _{C}(T)\right] }\right\} \right)  \nonumber \\
&=&\frac{1}{2}\left\{ 1\pm \frac{\mbox{e}^{-2\left| w(T)\right| ^{2}}}{%
1+\cos \varphi \mbox{e}^{-2\left| \alpha \right| ^{2}}}\left[ \mbox{e}%
^{-2\left| \alpha \right| ^{2}\mbox{e}^{-\gamma t}}\cosh \left( 4\mbox{e}^{-%
\frac{\gamma T}{2}}\mbox {Re}\left[ \alpha w^{*}(T)\right] \right) \right.
\right.  \nonumber \\
&&\left. \left. +\cos \varphi \mbox{e}^{-2\left| \alpha \right| ^{2}\left( 1-%
\mbox{e}^{-\gamma T}\right) }\cos \left( 4\mbox{e}^{-\frac{\gamma T}{2}}%
\mbox
{Im}\left[ \alpha w^{*}(T)\right] \right) \right] \right\} ,  \label{g6}
\end{eqnarray}
where $\varphi =0$ ($\pi $) for the first atom detected in the state $\left|
g\right\rangle _{1}$ ($\left| e\right\rangle _{1}$) and the signal $+(-)$
for the second atom detected in the state $\left| g\right\rangle _{2}$ ($%
\left| e\right\rangle _{2}$). Analyzing Eq. (\ref{g6}) one verifies that
if the second atom is detected instantaneously after the first one, $\gamma
T\ll 1,$ one gets
\begin{equation}
P_{{g \choose {e}}
}=\frac{1}{2}\left[ 1\pm \frac{\mbox{e}^{-2\left| \alpha \right| ^{2}}+\cos
\varphi }{1+\cos \varphi \mbox{e}^{-2\left| \alpha \right| ^{2}}}\right] ,
\label{g7}
\end{equation}
and for $\left| \alpha \right| \gg 1,$
\begin{equation}
P_{{g \choose {e}}}=\frac{1}{2}\left[ 1\pm \cos \varphi \right] ,  \label{g8}
\end{equation}
which is the result obtained in \cite{davidovich} without pumping: If the
first atom is detected in $\left| g\right\rangle $ or $\left| e\right\rangle
$, the field in $C$ collapses to an even or odd cat field state, $\left|
\Psi _{C}\right\rangle =\frac{1}{N}\left( \left| \alpha \right\rangle +\cos
\varphi \left| -\alpha \right\rangle \right) $, $\varphi =0$ or $\pi $
respectively.

Now let us suppose that the first atom is detected in the state $\left|
e\right\rangle $, then the odd cat state is generated in $C$. For $T\ll
t_{d} $ (the time interval between sequentially emitted atoms being quite
small) the second atom can be detected either in the state $\left|
g\right\rangle $, with conditional probability $P(e,g,T)\gtrsim 0$, or in
the state $\left| e\right\rangle $, with conditional probability $%
P(e,e,T)\lesssim 1$, and so on for the subsequent atoms crossing the
apparatus. In this manner the atoms crossing the apparatus sustain
(approximately) the superposition state. The measurement of the field state
in $C$ by the atoms refresh its superposition character, so turning the
environment induced decoherence almost ineffective. Thus if an experiment
can be done where $T\ll t_{d}$, a kind of Zeno effect takes place in a
continuous measurement process.

When the second atom is detected in a different state from the former, the
original cat state changes its parity. If one wishes to maintain the parity of
the original cat state a resonant interaction could be used to restore the
state of the field to its initial state. Such a process, could be outlined as
the feedback process reported in \cite{milburn}, once the resonant interaction
time can be controlled to produce a single photon exchange between the atom and
field. When the cavity field looses a photon
the state of the field flips from odd to even cat state and {\it vice-versa}%
. As the initial field state (prepared by the first atom) is an odd cat
state and the second atom is detected at $\left| g\right\rangle $, then a
conditional feedback process must be activated and the field flips to the
even cat state.

The case $T\gtrsim t_d$ is better understood by observing the behavior of
the conditional probabilities $P(g,e,T)$ and $P(e,e,T)$ in Figs. 5 and 6,
where these quantities are plotted as function of $\gamma T$ for several
values of pumping field intensity, $\left| F\right| ^2$, and for $\left|
\alpha \right| ^2=5$. In the absence of pumping both conditional
probabilities go to zero asymptotically because the field in $C$ ends in a
vacuum state. However, the pumping modifies this trend: the higher the
pumping field intensity the faster $P(g,e,T)$ and $P(e,e,T)$ will attain $%
1/2 $, an upper limit that does not depend on the intensity of the pumping.
This limit means that the pumping action drives the cavity field to a coherent
state and with the next atomic interaction another superposition state is
generated. It has 50\% of probability to be an even or odd cat state depending
on the state in which the second atom is measured. This process that guarantee
a 50\% efficiency for the detection of the same initial superposition is not
very useful if we do not introduce a supplementary process. The efficiency can
be increased if a conditional measurement is used for assuring that for each
`wrong' result (the atom not being detected in the required state) a resonant
feedback atom is sent through the cavity to flip the parity of the field state.
It is worth to mention that this process which guarantees an efficiency for the
generation of the same superposition state up to 93 \% was proposed in \cite
{cnot} for controlling the parity of a field cat state in a quantum logic gate
encoding.

It is important to note that the classical pumping acts on the cavity-field
relaxation time. The stronger the pumping intensity, $|F|^2$, the faster will
be the relaxation of any initial state to a coherent state. For $|F|^2=1$, the
time delay between sequentially emitted atoms should be about $\gamma T\gtrsim
3$, defining a minimum time interval for state reconstruction. While the
feedback process \cite{milburn,cnot} is fully dependent on the atomic detectors
efficiency, the proposed process for delaying the cavity-field decoherence does
not depend. Thus, this process is feasible as soon as each atom of the sequence
is prepared in the required state and time, as discussed above. Actually,
nowadays it is not an easy task to achieve an efficient control of atomic
injection for sending exactly one atom at a time in the cavity \cite{vitali}.
For instance, sending a single atom into a cavity means to send an atomic pulse
with an average number of $0.2$ atoms, making negligible the chance of finding
simultaneously two atoms in the cavity \cite{hagley} . However, the required
technology for energy supply - feeding the cavity continuously with a classical
source - is already available since it is employed in current experiments
\cite{raibruhaprl,brune}.
%
\section{Summary and discussion}
%
The proposed scheme of the paper shows how a classical pumping field drives
any
initial state prepared in a lossy cavity into a stationary coherent state.
The pumping compensates the lost energy due to the cavity damping mechanism;%
however, due to the phase insensitivity, this energy feeding does not
re-establish the initial superposition of two coherent states, destroyed during
the decoherence process. The pumping does not change the time of decoherence of
an initial cat state, which remains the same as in the free decoherence case,
showing that the information flows from the cavity field to the environment at
the same rate independently from the amount of supplied energy. However, the
combined action of pumping together with a sequential injection of atoms
interacting dispersively with the cavity field (atomic quantum non-demolition
measurement) can be used for partially conserving an initial cat state in the
cavity. This state can be partially conserved by an atom `measuring' the cavity
field state, thus re-establishing partially its original coherence.This result
is to be compared with that in \cite {davidovich}, where the mechanism of
atomic quantum non-demolition measurement is used without pumping the cavity.
In Figs. 5 and 6 we show that for large enough delay times between sequentially
injected atoms the action of pumping $(F\neq 0)$ contributes to reset the
initial cat state. This may be important in a practical implementation of
quantum processors.

The importance for seeking a process that may sustain the coherence of a
superposition state is based on the possibility of encoding information in
the field state. We expect that even and odd cat states could be used for
this purpose because they constitute an orthogonal basis, which should be a
sufficient condition to encode qubits. As reported in \cite{munro}, we can
consider the even cat state as being the $0$ qubit and the odd cat state as
the $1$ qubit,
\[
\left| 0\right\rangle _L\equiv \frac 1{N_{+}}\left( \left| \alpha
\right\rangle +\left| -\alpha \right\rangle \right) \quad \quad {\rm and}%
\qquad \left| 1\right\rangle _L=\frac 1{N_{-}}\left( \left| \alpha
\right\rangle -\left| -\alpha \right\rangle \right) .
\]
These states can only be used to encode qubits while as pure states,
however, dissipation precludes their existence as such. In conclusion, drawing
strategies to suppress or at least to delay the decoherence time is
therefore extremely important for technological purposes and worth to be
pursued.
%
\acknowledgments{MCO acknowledges the financial support from FAPESP
(Brazil). MHYM and SSM acknowledge partial support from CNPq (Brazil).}
%
%
\appendix
%
\section{Solution of the Heisenberg equation}
%
The solution to Eq. (\ref{b8}), goes closely along the lines of Louisell
\cite{louisell}, its Laplace transform is
\begin{eqnarray}
{\cal L}(\dot{A}) &\equiv &\int_0^\infty \mbox{e}^{-st}\dot{A}%
dt=-i\sum_k\lambda _kb_k(0)\int_0^\infty \mbox{e}^{-st}\mbox{e}^{-i(\omega
_k-\omega _0)t}dt  \nonumber \\
&&-\sum_k\left| \lambda _k\right| ^2\int_0^\infty \mbox{e}%
^{-st}dt\int_0^tA(t^{\prime })\mbox{e}^{i(\omega _k-\omega _0)(t^{\prime
}-t)}dt^{\prime }-iF\int_0^\infty \mbox{e}^{-st}\mbox{e}^{-i(\omega -\omega
_0)t}dt.  \label{i2}
\end{eqnarray}
The integrals give
\begin{equation}
\int_0^\infty \mbox{e}^{-st}dt\int_0^tA(t^{\prime })\mbox{e}^{i(\omega
_k-\omega _0)(t^{\prime }-t)}dt^{\prime }=\frac{\tilde{A}(s)}{s+i(\omega
_k-\omega _0)},  \label{i3}
\end{equation}
\begin{equation}
\int_0^\infty \mbox{e}^{-st}\mbox{e}^{-i(\omega _k-\omega _0)t}dt=\frac 1{%
s+i(\omega _k-\omega _0)},  \label{i4}
\end{equation}
\begin{equation}
\int_0^\infty \mbox{e}^{-st}\mbox{e}^{-i(\omega -\omega _0)t}dt=\frac 1{%
s+i(\omega -\omega _0)}\,,  \label{i5}
\end{equation}
and
\begin{equation}
\int_0^\infty \mbox{e}^{-st}\frac d{dt}\left[ A(t)\,\right] \,dt=s\tilde{A}%
(s)-A(0),  \label{i6}
\end{equation}
with $\tilde{A}(s)\equiv {\cal L}(A(t))$. Substituting these in Eq. (\ref{i2}%
), after a little algebra one gets
\begin{equation}
\tilde{A}(s)=\frac{A(0)-i\frac F{s+i(\omega -\omega _0)}-i\sum_k\frac{%
\lambda _kb_k(0)}{s+i(\omega _k-\omega _0)}}{\left[ s+\sum_k\frac{\left|
\lambda _k\right| ^2}{s+i(\omega _k-\omega _0)}\right] }.  \label{i7}
\end{equation}
The Wigner-Weisskopf approximation \cite{louisell} assumes that in the
denominator of the LHS in the above equation the frequency spectrum of the
reservoir is densely distributed around the cavity characteristic frequency $%
\omega _0$, such that one can replace the discrete sum by an integration
over the reservoir frequencies having a distribution $g(\omega )$ and do the
so-called `pole approximation',
\begin{eqnarray}
\sum_k\frac{\left| \lambda _k\right| ^2}{s+i\left( \omega _k-\omega
_0\right) } &=&-i\sum_k\frac{\left| \lambda _k\right| ^2}{\left( \omega
_k-\omega _0\right) -is}  \nonumber \\
&=&\lim_{s\rightarrow 0}\left\{ -i\int_0^\infty d\omega ^{\prime }\,\frac{%
g(\omega ^{\prime })\left| \lambda (\omega ^{\prime })\right| ^2}{\left(
\omega ^{\prime }-\omega _0\right) -is}\right\} \,.  \label{i8}
\end{eqnarray}
Considering only the first order shift in the simple pole in $\omega _0$ in
the above integral we have the Wigner-Weisskopf approximation for $%
s\rightarrow 0$
\begin{eqnarray}
\sum_k\frac{\left| \lambda _k\right| ^2}{s+i\left( \omega _k-\omega
_0\right) } &=&-i\int d\omega ^{\prime }\,g(\omega ^{\prime })\left| \lambda
(\omega ^{\prime })\right| ^2\left[ \frac 1{\left( \omega ^{\prime }-\omega
_0\right) }+i\pi \delta (\omega ^{\prime }-\omega _0)\right]  \nonumber \\
&=&\frac \gamma 2+i\Delta \omega ,  \label{i9}
\end{eqnarray}
where
\begin{equation}
\gamma \equiv 2\pi g(\omega _0)\left| \lambda (\omega _0)\right| ^2,
\label{i10}
\end{equation}
is the damping constant and
\begin{equation}
\Delta \omega \equiv -\int d\omega ^{\prime }\,\frac{g(\omega ^{\prime
})\left| \lambda (\omega ^{\prime })\right| ^2}{\omega ^{\prime }-\omega _0}%
\,,  \label{i11}
\end{equation}
is the frequency shift. So Eq.(\ref{i7}) can be written as
\begin{eqnarray}
\tilde{A}(s) &=&\frac 1{s+\frac \gamma 2+i\Delta \omega }A(0)-i\sum_k\frac{%
\lambda _k}{\left[ s+i(\omega _k-\omega _0)\right] \left( s+\frac \gamma 2%
+i\Delta \omega \right) }b_k(0)  \nonumber \\
&&-i\frac F{\left[ s+i\left( \omega -\omega _0\right) \right] \left( s+\frac %
\gamma 2+i\Delta \omega \right) }.  \label{i12}
\end{eqnarray}
Now the calculation of the inverse Laplace transform
\begin{eqnarray}
A(t) &=&\frac{A(0)}{2\pi i}\oint \mbox{e}^{st}\frac 1{s+\frac \gamma 2%
+i\Delta \omega }ds  \nonumber \\
&&-\frac 1{2\pi }\sum_k\lambda _kb_k(0)\oint \mbox{e}^{st}\frac 1{\left[
s+i(\omega _k-\omega _0)\right] }\frac 1{\left( s+\frac \gamma 2+i\Delta
\omega \right) }ds  \nonumber \\
&&-\frac 1{2\pi }F\oint \mbox{e}^{st}\frac 1{\left[ s+i\left( \omega -\omega
_0\right) \right] }\frac 1{\left( s+\frac \gamma 2+i\Delta \omega \right) }%
ds,  \label{i13}
\end{eqnarray}
where $A(t)=e^{-i\omega _0t}a(t)$ and disregarding the small frequency shift
$\Delta \omega $, gives after a little algebra the solution to the
Heisenberg equation (\ref{b3}),
\begin{equation}
a(t)=u(t)a(0)+\sum_kv_k(t)b_k(0)+w(t),  \label{i14a}
\end{equation}
where
\begin{equation}
u(t)=\mbox{e}^{-\frac \gamma 2t}\mbox{e}^{-i\omega _0t},  \label{i14}
\end{equation}
\begin{equation}
v_k(t)=-\lambda _k\mbox{e}^{-i\omega _kt}\frac{\left[ 1-\mbox{e}^{-\frac %
\gamma 2t}\mbox{e}^{i\left( \omega _k-\omega _0\right) t}\right] }{\omega
_0-\omega _k-i\frac \gamma 2},  \label{i15}
\end{equation}
and
\begin{equation}
w(t)=F\mbox{e}^{-i\omega t}\frac{\left[ 1-\mbox{e}^{-\frac \gamma 2t}\mbox{e}%
^{i\left( \omega -\omega _0\right) t}\right] }{\omega -\omega _0+i\frac %
\gamma 2}.  \label{i16}
\end{equation}
%
%

\newpage {\bf Figure Captions}

\bigskip

{\bf Fig. 1} Wigner distribution function for the superposition state $%
\left( \left| \alpha \right\rangle +\left| -\alpha \right\rangle \right) /%
\sqrt{2}$ with $\left| \alpha \right| ^{2}=5$. The central structure
represents the coherence of the quantum state.

\bigskip

{\bf Fig. 2} Wigner distribution function for the state of Fig. 1 evolved to
$\gamma t=1$. The original coherence was suppressed by the environment
action and the state suffers a continuous displacement due to the pumping
field.

\bigskip

{\bf Fig.3} Asymptotic Wigner distribution function for the state of Fig. 1.
The original superposition state evolved asymptotically to a coherent state
due to the classical pumping.

\bigskip

{\bf Fig. 4} The evolution in time (in units of $\gamma ^{-1}$) of the
linear entropy for the continuously pumped initial superposition state. The
pumping does not affect the coherence terms, the state evolves
from a pure state to a mixture and then to a pure state again, as in the
absence of the pumping, but the final state is a coherent state
instead of a vacuum state.

\bigskip

{\bf Fig. 5} The conditional probability $P(g,e,T)$ (first atom in $\left|
e\right\rangle $ and second in $\left| g\right\rangle $) increases with the
interaction time $T$ (in units of $\gamma ^{-1}$) as the pumping field
intensity increases, saturating at $0.5$.

\bigskip

{\bf Fig. 6} As like as $P(g,e,T)$ in Fig. 5, the conditional probability $%
P(e,e,T)$ (first and second atoms in $\left| e\right\rangle $ ) increases with
the interaction time $T$ as the pumping field intensity increases, saturating
at $0.5$. This means that the cavity field state has $50\%$ chance to be left
in an even or odd cat state.
\end{document}